# Counterintuitive inverse superconducting transition beyond $^4$He-cooling limit

Haowen Han[1†], Yi Bian[1†], Tong Ma[1†], Yusong Zhao[1†], Nuofu Chen[2]*, Chuanying Xi[3], Ze Wang[3], Binghui Ge[4], Hongliang Dong[5], Jia-Cai Nie[6,7]*, Ho-Kwang Mao[5,8], and Jikun Chen[1]*

[1] *School of Materials Science and Engineering, University of Science and Technology Beijing, Beijing 100083, China.*

[2] *School of Renewable Energy, North China Electric Power University, Beijing 102206, China.*

[3] *Anhui Key Laboratory of Low-Energy Quantum Materials and Devices, High Magnetic Field Laboratory, HFIPS, Chinese Academy of Sciences, Hefei, Anhui 230031, China.*

[4] *Institute of Physical Science and Information Technology, Anhui University, Anhui 230601, China.*

[5] *Center for High Pressure Science and Technology Advanced Research, Shanghai 201203, China.*

[6] *School of Physics and Astronomy, Beijing Normal University, Beijing 100875, China.*

[7] *Key Laboratory of Multiscale Spin Physics, Ministry of Education, Beijing Normal University, Beijing 100875, People's Republic of China.*

[8] *Shanghai Key Laboratory of Material Frontiers Research in Extreme Environments (MFree), Institute for Shanghai Advanced Research in Physical Sciences (SHARPS), Shanghai 201203, China.*

[†] These authors contributed equally to this work

**Abstract**

Thermally driven quantum-orders observed in exceptional instances—e.g., prethermal quantum states[1,2], thermal quantum Hall-effect[3,4], thermally driven quantum refrigerator[5] and magnetism-coupled superconductivity[6-9]—may redefine the role of thermal-fluctuation from a source of decoherence to a resource for coherent-state engineering. While preliminary signs of counterintuitive temperature-rise-triggered superconductivity manifested in $CeCu_2Si_2$[6], $ErRh_4B_4$[7], $Ho_{1.2}Mo_6S_8$[8] and (La,Ce)$Al_2$[9], their critical-temperatures ($T_{c\text{-inv}}$) remain below Kelvin-range, precluding substantial applications. Here, we report field-modulated inverse-superconducting-transitions above $^4$He-cooling-limit in Eu-based infinite-layer nickelates ($Eu_xNd_{1-x}NiO_2$ and $Eu_xPr_{1-x}NiO_2$) grown on a substrate under both overdoped and underdoped regimes. Paradigmatically, superconductivity with zero-resistance is confined between $T_{c\text{-inv}}$ (~2.6-5.4 K) and another higher normal-$T_c$, rising and decreasing with applied magnetic-field, respectively. Starting from the resistive-state below $T_{c\text{-inv}}$, the inverse-superconducting-transition is driven by not only temperature-rising, but also current-density, while superconductivity further vanishes at higher temperature and current thresholds. The Kelvin-range inverse superconducting transition is plausibly explained by temperature-induced alternating dominance of effective magnetic-fields—arising from $Eu^{2+}4f^7$-related compensations[10-13]—relative to the upper-critical-field[12,14-17]. Furthermore, an extended-phenomenological-framework is also supported[18] by reemerged superconductivity below ~300 mK under magnetic-field, giving rise to an unprecedented temperature-induced reentrant superconductivity. Our findings establish magnetic-interaction-reconfigured high-$T_c$ systems as fertile platforms for exploring quantum phenomena that reverse thermal-decoherence paradigm, also enabling antithetical-designs to unlock untapped application-scenarios for quantum-phase-transition devices[19-23].

## 1. Introduction

Thermal fluctuations constitute the primary decoherence source disrupting macroscopic quantum ordering—an effect epitomized by superconductivity[24,25], which overwhelmingly emerges upon cooling down below a critical temperature ($T_c$). Notwithstanding this consensus, its counterintuitive reversal—superconductivity induced by temperature rise beyond a second and lower critical threshold—was predicted by Jaccarino and Peter in 1962[14], originating from the internal exchange-field ($H_J$) compensated by the externally applied magnetic-field ($H$). In 1972, Fischer further established an improved Jaccarino-Peter (J-P) compensation superconducting phase-diagram, displaying dichotomous superconducting domes with nonmonotonic boundaries along $H$ and $T$, respectively, giving rise to both high-field and low-field superconductivity[15]. The former primarily counts for the magnetic-field reentrant superconducting behavior[14,15], as has been observed in multiple material systems, such as Eu-containing Chevrel phase compounds[16,17], λ-$(BETS)_2FeCl_4$-based organic conductors[26], and uranium-based ferromagnetic/heavy-fermion superconductors[27]. More recently, the field-reentrant superconductivity likely via J-P compensation effect was also observed in a high-$T_c$ system of Eu-doped infinite layer (IL) nickelates[10-13], displaying $T_c$ up to 40 K[10,28]. Conversely, the latter yet remains unsubstantiated in these systems[16,17,27,29], otherwise applying a low-magnitude $H$ to traverse the low-field superconducting dome within the J-P phase-diagram would confine superconductivity between two critical temperatures[14,15], $T_{c\text{-inv}}$ and $T_c$. Given such premise, temperature rising across $T_{c\text{-inv}}$ will trigger counterintuitive inverse superconducting transitions. Nevertheless, experimental observations contradict the J-P theoretical predictions, since $H$ and $T$ both approaching to zero always results in merged low-field and high-field superconducting regions in J-P phase-diagram of existing systems[16,17,26,27]. The tentative cause lies in the robust upper critical field ($H_{c2}$) at low temperatures compared to a saturating $H_J$, posing formidable challenge to eliminate superconductivity anymore under low magnetic-fields. In early reports, analogous superconductivity confined between $T_{c\text{-inv}}$ and $T_c$, was preliminarily observed in heavy-fermion $CeCu_2Si_2$ ($T_{c\text{-inv}}$~0.1 K, $T_c$~0.2 K)[6], long-range magnetic ordering coupled $ErRh_4B_4$ ($T_{c\text{-inv}}$~0.9 K, $T_c$~8.7 K)[7] and $Ho_{1.2}Mo_6S_8$ ($T_{c\text{-inv}}$~0.6 K, $T_c$~1.3 K)[8], and Kondo-superconductor $(La,Ce)Al_2$[9] ($T_{c\text{-inv}}$~0.2 K, $T_c$~1.3 K) under magnetic-fields. Nevertheless, their underlying mechanisms likely lie beyond J-P compensation and are associated with the Kondo-effect[9,30], since the absence of magnetic-field reentrant superconductivity in these systems[6,7] deviates from the high-field prediction of the J-P phase-diagram[14,15]. To date, the experimentally detected inverse superconducting transitions have yet remained below the liquid $^4$He-evaporative cooling-limit[6,7], precluding substantial applications owing to cryogenic constraints.

Bridging magnetic exchange-field interactions with high-$T_c$ superconducting systems may establish a paradigmatic J-P compensation platform for elevated $T_{c\text{-inv}}$ and also new quantum phenomena discoveries. Should a $T_{c\text{-inv}}$ beyond the $^4$He-cooling-limit be realized, it would reshape the fundamental architecture in designing applicable quantum devices, since the thermal and/or electrical fluctuations in such frameworks are no longer the destroyer but the activator of hidden quantum orders. Notably, the reversal of thermal decoherence paradigm is expected to enable antithetical designs for existing quantum devices based on superconducting transitions, such as transition-edge sensors[19,20], superconducting nanowire single-photon detectors[21] and superconducting nanowire cryotron switches[22]. Compared to cuprates intertwined with antiferromagnetic ordering[31-34], a better candidate would be the emerging Eu-doped IL-nickelates, manifesting substantial magnetic instabilities and interactions with unconventional superconductivity from the half-filled $Eu^{2+}$-$4f^7$[12,13,35]. Their resultant effective magnetic-field ($H_{tot}$=$H$+$H_J$) acting on Cooper pairs competes with $H_{c2}$, giving rise to exotic magnetic-field induced reentrancy[10-13]. Nevertheless, a further obstacle lies in their more robust $H_{c2}$ in high-$T_c$ systems at low temperatures[12,36,37] compared to conventional superconductors, and this increase the difficulty to eliminate their superconductivity via low-field magnetic interactions. Overturning $H_{tot}$ compared to the robust $H_{c2}$ at both low temperatures and low magnetic-fields may require unlocking new freedoms to push the limits of magnetic interactions with superconductivity for IL-nickelates.

In this work, we report an inverse superconducting transition beyond the $^4$He-cooling-limit achieved in Eu-doped IL-nickelates grown on a special substrate not the same as our previous work[10], driven by both temperature-rise and current-density under Tesla-range magnetic-fields. As demonstrated in several compositions of $Eu_xNd_{1-x}NiO_2$ and $Eu_xPr_{1-x}NiO_2$, elevating the temperature across $T_{c\text{-}inv}$ of ~2.6-5.4 K drives the samples into superconducting state with zero-resistance, which further vanished at a higher normal-$T_c$. In a similar vein, the inverse superconducting transition starting from a resistive state below $T_{c\text{-}inv}$ has been also driven by current-density across a lower threshold, while further enlarging current-density results in the vanished superconductivity. Systematically modulating Eu-compositions to establish a complete superconducting phase-diagram unveils that the inverse superconducting transition behavior emerges at both overdoped and underdoped boundaries of the superconducting dome, but is shielded under optimum-doped regime. The temperature-induced reciprocal fluctuations in primacy between $H_J$ compensation akin to J-P effect and $H_{c2}$ plausibly explains the inverse superconducting transition emerging at Kelvin temperature range. Nevertheless, the reemerged superconductivity upon cooling into $10^1$-$10^2$ mK also suggests extension beyond the J-P compensation framework, giving rise to more intriguing and unprecedented temperature-induced reentrant superconducting phenomena. Our findings establish magnetic-interaction-reconfigured high-$T_c$ systems as fertile platforms for exploring quantum phenomena reversing thermal decoherence paradigm, also expanding application domains of quantum-transition devices via complementary designs.

## 2. Results and discussions

### Counterintuitive inverse superconducting transition driven by temperature-rise

The characteristic inverse superconducting transition behavior is demonstrated in Fig. 1a and 1b, measured from the temperature-dependent resistivity ($\rho$-$T$) curves under different magnetic-fields ($H$) for $Eu_xNd_{1-x}NiO_2$. With an increase in $H$ from zero, a bulge emerges in $\rho$-$T$ from the low temperature-range (e.g., 2-7 K) and evolves towards the top-right quadrant until $H$=4 T, afterwards fading out. By further increasing $H$ (see also Extended Data Fig. 1), the superconducting transition slope in $\rho$-$T$ curves shifts towards higher temperatures enabling reentrant superconductivity until 20 T; after this point, the trend reverses as $H$ continues increasing up to 35 T. Of particular note is the superconductivity confined between two distinct critical temperatures, $T_{c\text{-inv}}$ and $T_c$, under $H$ within 1-2 T shown in Fig. 1b, which gives rise to superconductivity triggered not only by temperature decrease below $T_c$ but also by temperature increase beyond $T_{c\text{-inv}}$. Both inverse and normal superconducting transitions are confirmed to be reversible (see Fig. 1b inset) and path-independent with respect to $H$ (see Supplementary Information Fig. 5). Additionally, weak anisotropy of both normal and inverse superconducting transitions in our samples when altering the angle ($\theta$) between $H$ and the normal of the film surface (see Extended Data Fig. 2). Hence, similar inverse superconducting transition behaviors were observed for applied $H$ at both $\theta$=0º as well as $\theta$=90º.

Additional evidence for the inverse superconducting transition is provided by the cross-linking in magnetic-field dependent resistivity ($\rho$-$H$) curves measured at stepwise increasing temperatures, as shown by Fig. 1c. Similar to the previous reports on Eu-doped IL-nickelates[10-13], our sample also displays magnetic-field reentrant superconductivity: for example, the superconductivity vanished at a low-field $H_1$ (~1 T at 3 K), but reentered at a higher-field $H_2$ (~7 T at 3 K), and vanished again at the highest $H_3$ (~30 T at 3 K; see Extended Data Fig. 1c). Nevertheless, what sets the present field-reentrant behavior apart is that $H_1$ increases at higher temperatures, indicating temperature-rise-induced more robust pairing against magnetic-field within 2-6 K. Constructing a colour map of resistivity versus $H$ and $T$ ($\rho$-$H$-$T$) in Fig. 1d showcases an archetypal J-P compensation-modulated superconducting phase-diagram, containing both low-field and high-field superconducting regions. The inverse superconducting transition stems from the non-monotonic $H$-$T$ dependence framing the low-field superconducting phase along $T$, e.g., as indicated by the open circles corresponding to 0.5% of the normal-state resistivity (see Fig. 1d).

In a similar vein, the inverse superconducting transition should also be driven by increasing the electric current-density applied to a resistive state below $T_{c\text{-inv}}$, in addition to a temperature rise across $T_{c\text{-inv}}$. This is demonstrated in Extended Data Fig. 3 by measuring the current-voltage ($I$-$V$) characteristics at 2 K under magnetic-fields for $Eu_xNd_{1-x}NiO_2$. The current-driven transitions into superconducting states under 1-2 T are demonstrated by an abrupt drop of $V$ to zero across a lower current-threshold, while further increasing $I$ beyond a higher threshold abruptly elevates $V$, indicating elimination of superconductivity. Another counterintuitive point worth noting is that a larger current is required to eliminate the pristine superconductivity when the applied magnetic-field increases into a higher range (e.g., 4–9 T).

**Plausible explanation by temperature altered primacy between $H_{tot}$ and $H_{c2}$**

To delve into the underlying mechanism, we performed high-field magnetoelectric transport measurements on $Eu_xNd_{1-x}NiO_2$, with $H$ applied both parallel and perpendicular to the substrate normal, as shown in Extended Data Fig. 1a and 1b, respectively. Accordingly, their $\rho$-$H$-$T$ colour maps were constructed for higher ranges of magnetic-fields, as shown in Fig. 2a and Extended Data Fig. 4a for $\theta$=0º and $\theta$=90º, respectively. By fitting these high-field transport results with the modified Ginzburg-Landau (G-L)[25] and Fischer models[15], we obtained the saturated exchange-field ($H_{J0}$). At each given magnetic-field, the temperature dependence of the internal exchange-field ($H_J$-$T$) and further the effective magnetic-field ($H_{tot}$-$T$, $H_{tot}$=$H_J$+$H$) were established according to Brillouin function, while the temperature dependence of $H_{c2}$ was constructed by unmodified Werthamer-Helfand-Hohenberg (WHH) theory[38]. Plotting the $H_{tot}$-$T$ for typical $H$ (e.g., 0.5 T, 2 T, and 20 T) together with $H_{c2}$-$T$ (see Fig. 2b) unveils their two intersection points under $H$=1-2 T, corresponding to $T_{c\text{-}inv}$ and $T_c$, which confine the superconducting region with $H_{c2}$ exceeding $H_{tot}$ in between. In contrast, when $H$ is either increased above or decreased below the 1–2 T range, the suppressed $H_{tot}$ at low temperatures results in only one intersection point between $H_{tot}$-$T$ and $H_{c2}$-$T$ (see Fig. 2b), thus impeding the inverse superconducting transition.

**Eu-based phase-diagram for inverse superconducting transition**

Since the hole doping-level originating from the $Eu^{2+}$ partitions the superconducting dome of $Eu_xRE_{1-x}NiO_2$ into overdoped, optimum-doped and underdoped regions[10,39,35], the following vital issue to address is what Eu-content gives rise to the inverse superconducting transition. By tuning the Eu/Nd ratio in $Eu_xNd_{1-x}NiO_2$, we observed analogous inverse superconducting transitions and field-reentrant superconductivity in other overdoped compositions, such as $Eu_xNd_{1-x}NiO_2$ and $Eu_xNd_{1-x}NiO_2$ (see Extended Data Figs. 5 and 6, respectively). Zero resistance vanishes as the Eu-content increases further toward the overdoped boundary (see Extended Data Fig. 7). Conversely, the inverse superconducting transition also emerged in the underdoped $Eu_xNd_{1-x}NiO_2$, together with magnetic-field reentrant superconductivity, as shown in Fig. 3c (see $\rho$-$T$ for other underdoped samples in Extended Data Fig. 8). Nonetheless, the $T_{c\text{-}inv}$ is lower compared to the overdoped $Eu_xNd_{1-x}NiO_2$ (see the inset of Fig. 3c).

In stark contrast, the optimum-doped $Eu_xNd_{1-x}NiO_2$ exhibits no inverse superconducting transition or field-reentrant superconductivity, showing only conventional superconducting behavior. This is demonstrated by the $\rho$-$T$ data for $Eu_xNd_{1-x}NiO_2$ measured under $H$ applied along the film surface ($\theta$=90º) up to 35 T in Fig. 3a, and the inset shows its $\rho$-$H$ behavior measured at various temperatures below $T_c$. Similar results are also observed $H$ is applied along the surface normal ($\theta$=0º), as shown in Supplementary Fig. 6. The $\rho$-$T$ and $\rho$-$H$ curves measured for other optimum-doped compositions are further shown in Extended Data Fig. 9. Further construction of the $\rho$-$H$-$T$ colour map in Fig. 3b reveals a single continuous superconducting region with a monotonic $H$-$T$ dependence at 50% of the normal-state resistivity. Robust superconductivity at optimal doping is also confirmed by the large critical current-density ($J_c$) reaching 273 kA/cm$^2$ at 2 K (see Supplementary Fig. 7), which is comparable to that of other IL-nickelates[10,12,40].

Integrating the results above to construct a complete Eu-based superconducting phase-diagram in Fig. 3d shows that inverse superconducting transitions emerge at both the underdoped and overdoped boundaries, coinciding with the field-reentrant superconductivity. In contrast, robust superconductivity in the optimally doped region completely screens both the inverse superconducting transition and field-induced reentrant superconductivity. Apart from the $Eu_xNd_{1-x}NiO_2$ system, we also observed inverse superconducting transition behavior in the Eu-Pr system, e.g., $Eu_xPr_{1-x}NiO_2$, as demonstrated in Extended Data Fig. 10.

**Temperature-induced reentrant superconductivity extending beyond J-P framework**

Further extending the $\rho$-$T$ measurements to lower temperatures down to $10^1$-$10^2$ mK reveals an intriguing reemerged superconductivity below a third critical temperature ($T_c$') of ~100-300 mK under 0-3 T (see Supplementary Fig. 8), which was not observed in $CeCu_2Si_2$[6], $ErRh_4B_4$[7], $Ho_{1.2}Mo_6S_8$[8] and $(La,Ce)Al_2$[9]. This establishes a more exotic temperature-induced reentrant superconductivity, as shown in Fig. 4a for $Eu_xNd_{1-x}NiO_2$. Typically, as the temperature increases from 60 mK, the ground-state superconductivity is firstly destroyed when the temperature acrosses $T_{c,0}$'~130-150 mK, the reemerges at $T_{c\text{-inv},0}$~3.5-5.5 K, and is finally destroyed again at $T_{c,0}$~6.7-9.8 K. In contrast to magnetic-field reentrant superconductivity observed in multiple systems including $UTe_2$[41,42], Eu-containing Chevrel phase compounds[16,17], $\lambda$-$(BETS)_2FeCl_4$[26] and moiré graphene[43], and Eu-doped IL nickelates[12,13], such unprecedented temperature-induced reentrant superconductivity has scarcely been identified in other known superconducting systems, to the best of our knowledge.

A more complete low-field J-P compensation superconducting phase-diagram has been constructed, by integrating low-temperature results, as shown by the $\rho$-$H$-$T$ colour map in Fig. 4b. Particularly in the low-temperature range, the $H$-$T$ tendency framing the superconducting region clearly deviates from predictions combining J-P compensation and mean-field theory[15,38], suggesting an anomalous $H_{c2}$ upturn, as previously observed in IL-nickelates[36,44-46]. This is likely related to effects including $Eu^{2+}$ spin correlations[47], dissipation-driven phase transitions[48], vortex-related peak effects[49], or a quantum Griffiths singularity arising from quenched disorder[18]. We also summarized the relationships of $\mu_0H$-$T_{c,0}$', $\mu_0H$-$T_{c\text{-inv},0}$ and $\mu_0H$-$T_{c,0}$ as plotted in Fig. 4c, where an opposite $\mu_0H$-$T_{c\text{-inv},0}$ tendency is observed, in stark contrast to that of $\mu_0H$-$T_{c,0}$' and $\mu_0H$-$T_{c,0}$. This indicates a negative pairing strength associated with the inverse superconducting transition (e.g., -0.19 T/K), where pairing is promoted rather than destroyed by the applied magnetic-fields. A similar negative pairing strength is also observed for other compositions that exhibit inverse superconducting transitions, as shown Supplementary Fig. 9.

## 3. Conclusion

For the first time, we identified the field-modulated inverse superconducting transitions above the $^4$He-cooling limit in $Eu_x RE_{1-x}NiO_2$ grown on a special substrate differs to our previous work[10] at both overdoped and underdoped phase-boundaries of the superconducting dome. The inverse superconducting transitions are driven by a lower-threshold critical temperature $T_{c\text{-inv}}$ to reach robust zero resistance, which is then eliminated at normal critical temperature $T_c$. In a similar vein, superconductivity has also been triggered by current-density from a resistive state below $T_{c\text{-inv}}$, and further vanishes at a higher current-threshold. While the temperature-induced alternating dominance between effective-field and $H_{c2}$ provides a plausible explanation of inverse superconducting transition in the Kelvin range, reemerged superconductivity below sub-Kelvins reveals extended framework, giving rise to unprecedented temperature-induced reentrant superconducting phenomenon. These collective findings establish magnetic ordering intertwined high-$T_c$ systems as a fertile platform for exploring quantum phenomena that overcome thermal decoherence, thus redefining architectures to broaden the scope of quantum-phase-transition devices.

**Competing interests:** We declare no competing financial interest.

**Additional information:** Supplementary Information is available for this manuscript.

**Correspondences:** Correspondence should be addressed: Prof. Jikun Chen (jikunchen@ustb.edu.cn), Prof. Jia-Cai Nie (jcnie@bnu.edu.cn) and Prof. Nuofu Chen (nfchen@ncepu.edu.cn).

**Author contributions:** J.C. conceived the project, collaborating with J.N. and N.C.; H.H., Y.B., T.M. and Y.Z. contributed equally to this work. H.H., Y.B. and T.M. performed the transport measurements as well as the XRD and XPS measurements; Y.Z. synthesized the perovskite precursor films; H.H. and T.M. performed the soft chemical reductions. H.H., C.X., and Z.W. performed the high-field magneto-transport measurements, while H.H. performed the low temperature transport measurements. B.G. conducted the STEM experiments. H.H and Y.B performed curve fitting advised by J.C and J.N. N.C, H.D and K.M. provide further experimental supports and constructive discussions. All authors analysed and discussed the data. J.C. wrote the manuscript, assisted by H.H., J.N, Y.B and N.C, also with input from all authors.

**Acknowledgments:** This work was supported by the National Key Research and Development Program of China (No. 2021YFA0718900), the National Natural Science Foundation of China (No. 62474017, 92577108 and 12474001), the Natural Science Foundation of Guangdong Province of China (Grant No. 2025A1515011071) and the Guangdong Provincial Quantum Science Strategic Initiative (Grant No. GDZX2501006). We also thank the WM1 of the Steady High Magnetic Field Facility, Chinese Academy of Sciences, for the assistance on the experiment.

**Figures and Captions**

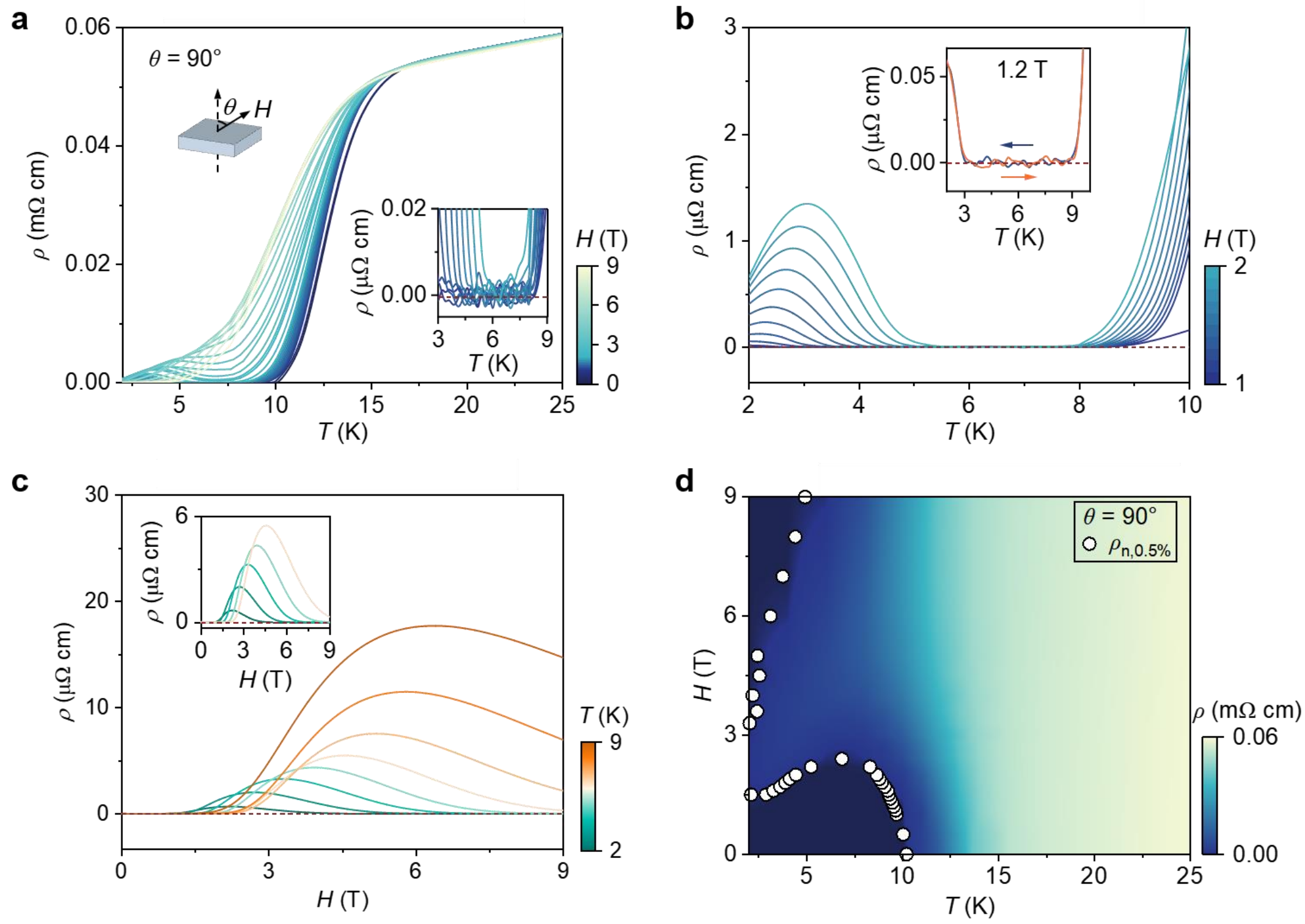

**Fig. 1 | Inverse superconducting transitions beyond $^4$He cooling-limit. a,** Temperature-dependent resistivity of overdoped $Eu_xNd_{1-x}NiO_2$ grown on a special substrate measured under magnetic-fields up to 9 T with $\theta = 90°$, where $\theta$ is the angle between the field and the substrate normal. Inset, detailed region showing zero-resistance. **b,** Magnified view of $\rho$-$T$ curves between 1 T and 2 T, showcasing a superconducting state confined between the inverse transition temperature $T_{c\text{-inv}}$ and the conventional transition temperature $T_c$. Inset, showcasing the reversibility of the inverse-/normal- superconducting-transitions. **c,** Field-dependent resistivity measured at selected temperatures, showing field-induced suppression and re-entrance of superconductivity. Inset, magnified view for low-field region. **d,** Resistivity-map plotted versus temperature and magnetic-field for $\theta$=90°. Open circles denote the 0.5%$\rho_n$ criterion.

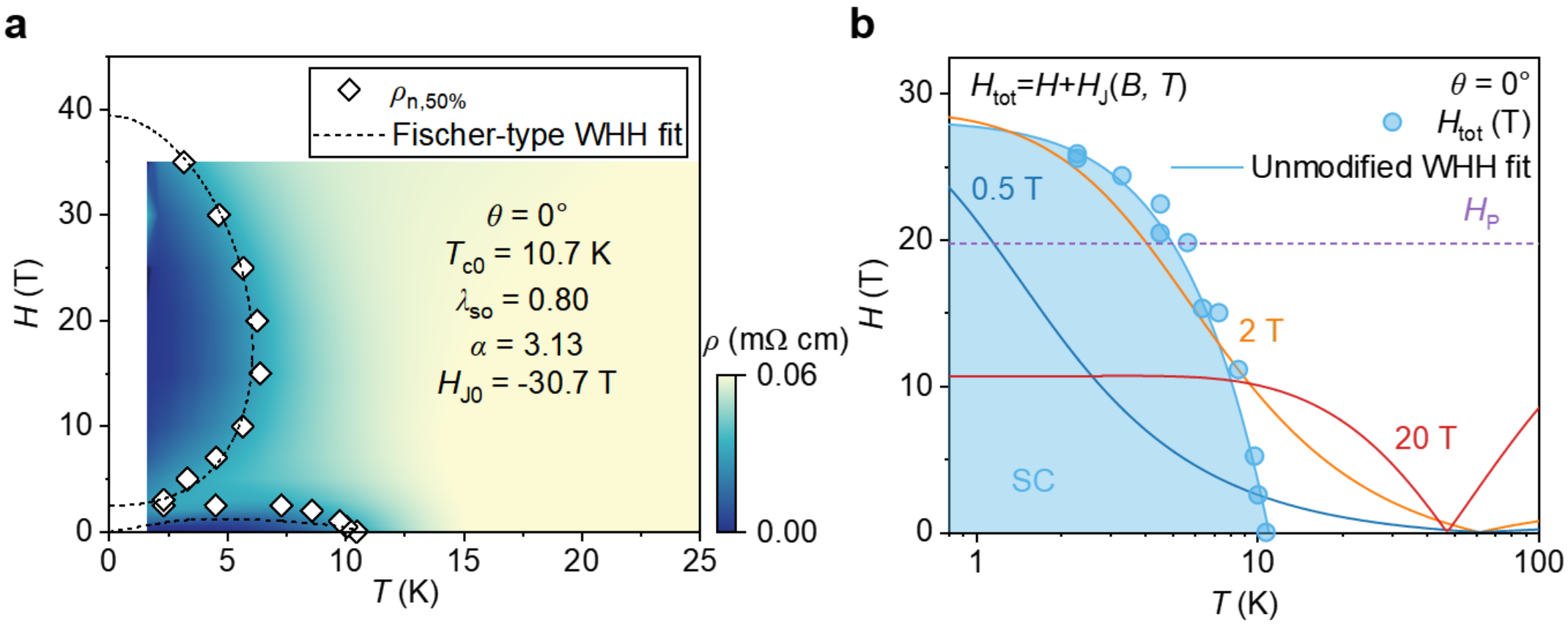

**Fig. 2 | Mechanism underlying Kelvin-ranged inverse superconducting transition. a,** High-field resistivity map of the overdoped $Eu_xNd_{1-x}NiO_2$ measured with the magnetic-field applied along the substrate normal ($\theta$=0°) up to 35 T. Open diamonds denote the superconducting boundary extracted using the 50%$\rho_n$ criterion; the dashed line shows the Fischer-type WHH fitting result[15,17]. **b,** Temperature dependence of the effective magnetic-field $H_{tot}=H+H_J(B, T)$ and the upper critical field $H_{c2}$(T) at representative applied fields of 0.5, 2 and 20 T. The intersections between $H_{tot}$(T) and $H_{c2}$(T) define the lower inverse-transition boundary $T_{c\text{-inv},0}$ and/or the higher conventional transition boundary $T_{c,0}$; the blue-shaded region marks the superconducting phase stabilized where $H_{c2}$ exceeds $H_{tot}$.

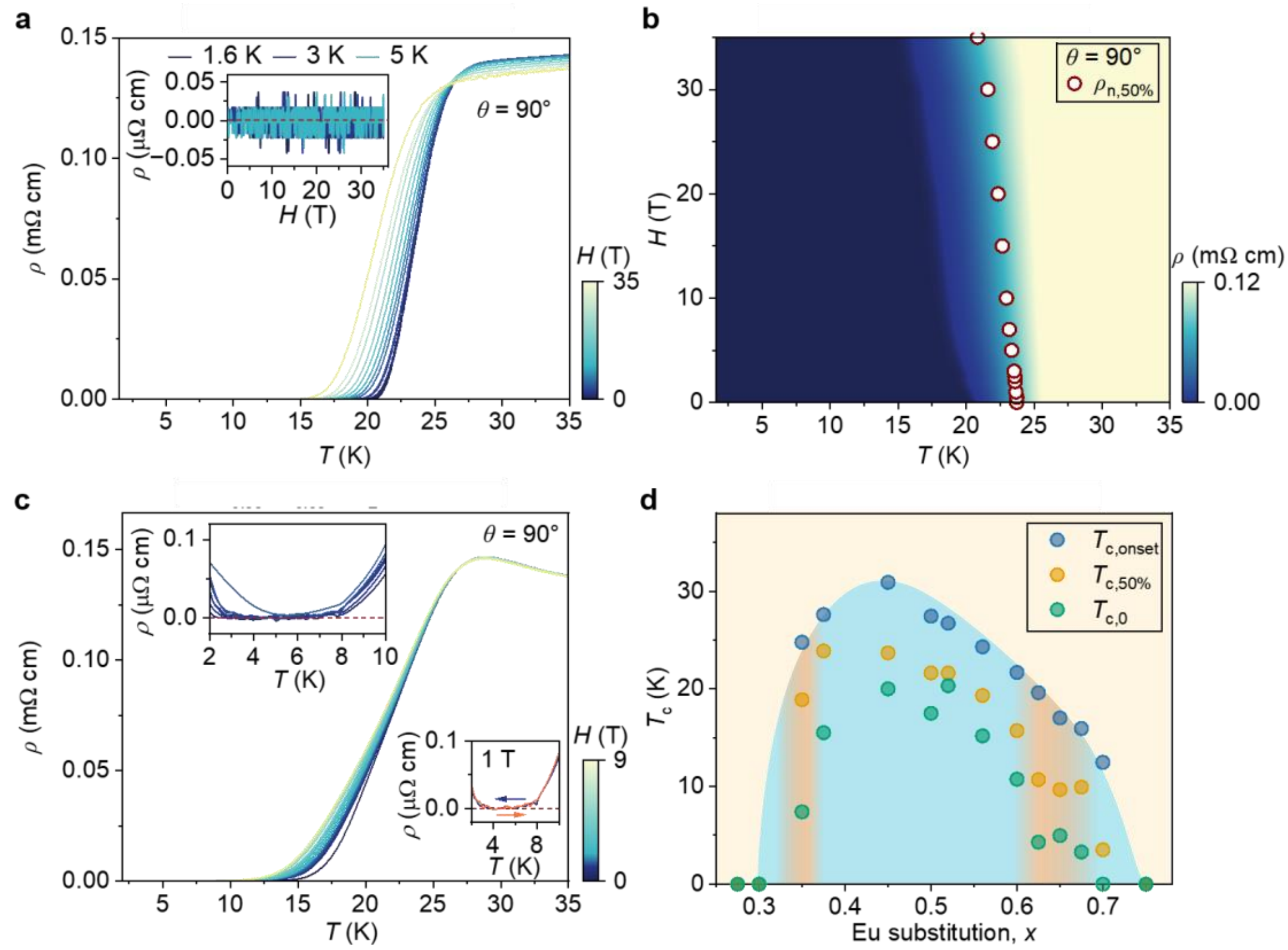

**Fig. 3 | Eu-based superconducting phase-diagram for inverse superconducting transitions.** **a,** Temperature-dependent resistivity of the optimum-doped $Eu_xNd_{1-x}NiO_2$ measured under magnetic-fields up to 35 T imparted perpendicular to the substrate normal ($\theta$=90°). Inset, field-dependent resistivity measured at low temperatures. **b,** Resistivity map of $Eu_xNd_{1-x}NiO_2$ as a function of temperature and magnetic-field. Open circles denote the 50%$\rho_n$ criterion, indicating a single and monotonic superconducting boundary. **c,** Temperature-dependent resistivity of the underdoped $Eu_xNd_{1-x}NiO_2$ measured under magnetic-fields up to 9 T with $\theta$=90°. Insets, magnified view of inverse superconducting transitions, highlighting the zero-resistance (upper-left inset) and reversibility (bottom-right inset). **d,** Superconducting phase-diagram of $Eu_xNd_{1-x}NiO_2$ as a function of Eu substitution. Symbols denote $T_{c,onset}$, $T_{c,50\%}$ and $T_{c,0}$. The orange-shaded regions mark the Eu-compositions giving rise to the inverse superconducting transitions.

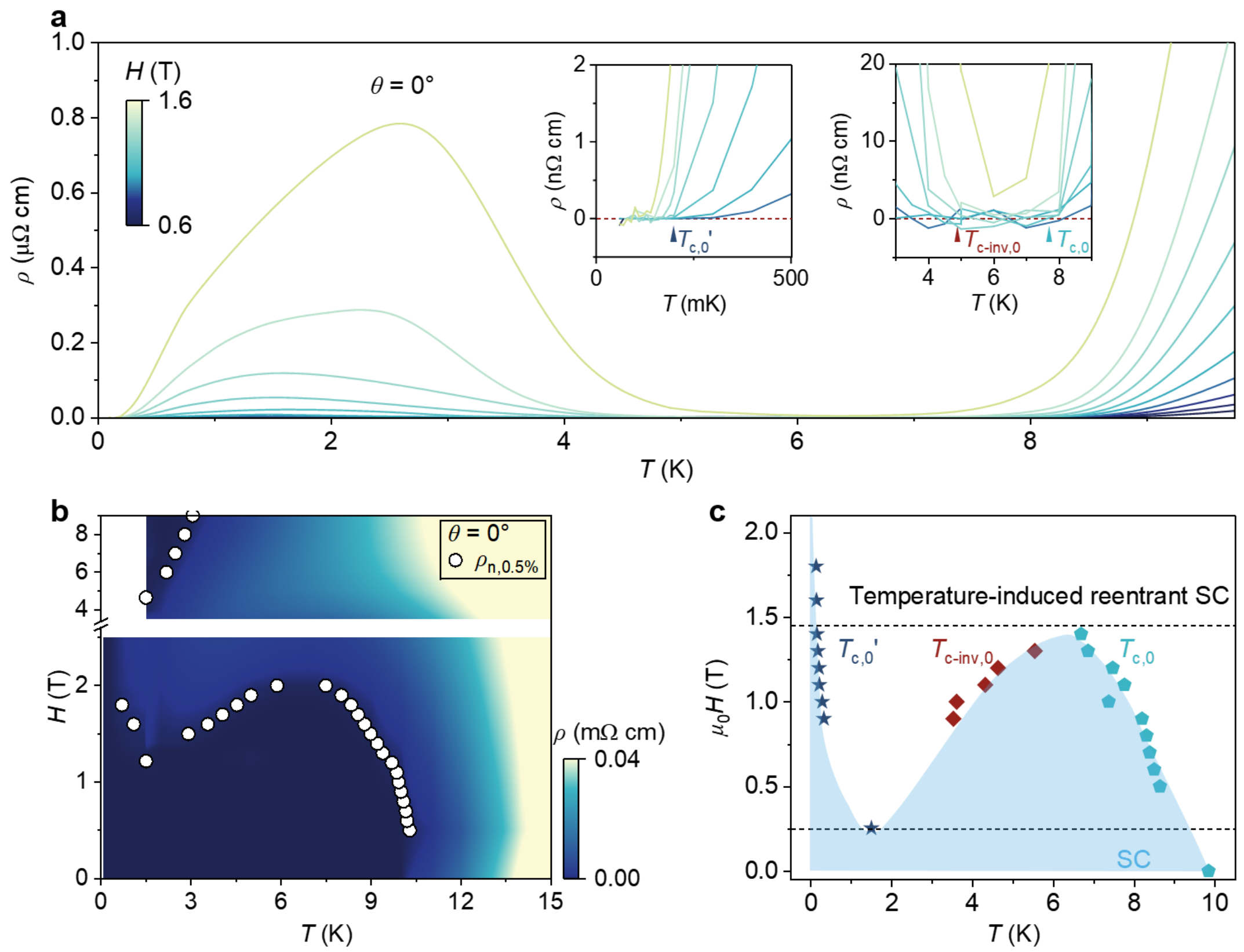

**Fig. 4 | Extended framework enabling temperature-induced reentrant superconductivity.** **a,** Temperature-dependent resistivity of overdoped $Eu_xNd_{1-x}NiO_2$ further integrating the low temperature range (e.g., 60 mK-2 K) measured under magnetic-fields. Left inset shows the first superconducting quench with temperature rise across $T_{c,0}'$ (left) at sub-kelvin range. Right inset shows temperature-rise induced reentrant superconductivity across $T_{c\text{-inv},0}$ and afterwards second quench again at a higher $T_{c,0}$. **b,** Resistivity map as a function of temperature and magnetic-field, with extension to the sub-kelvin regime. The open circles denote the $0.5\%\rho_n$ criterion. Deviations from the Jaccarino-Peter compensation framework are observed by the reemerged superconductivity at low temperature and low field region. **c,** $\mu_0H$-$T$ diagram constructed from $T_{c,0}'$, $T_{c\text{-inv},0}$ and $T_{c,0}$. The blue-shaded region denotes the superconducting phase, while the temperature-induced reentrant superconductivity is expected to emerge between the two critical magnitudes of magnetic-field as marked by the dashed lines.

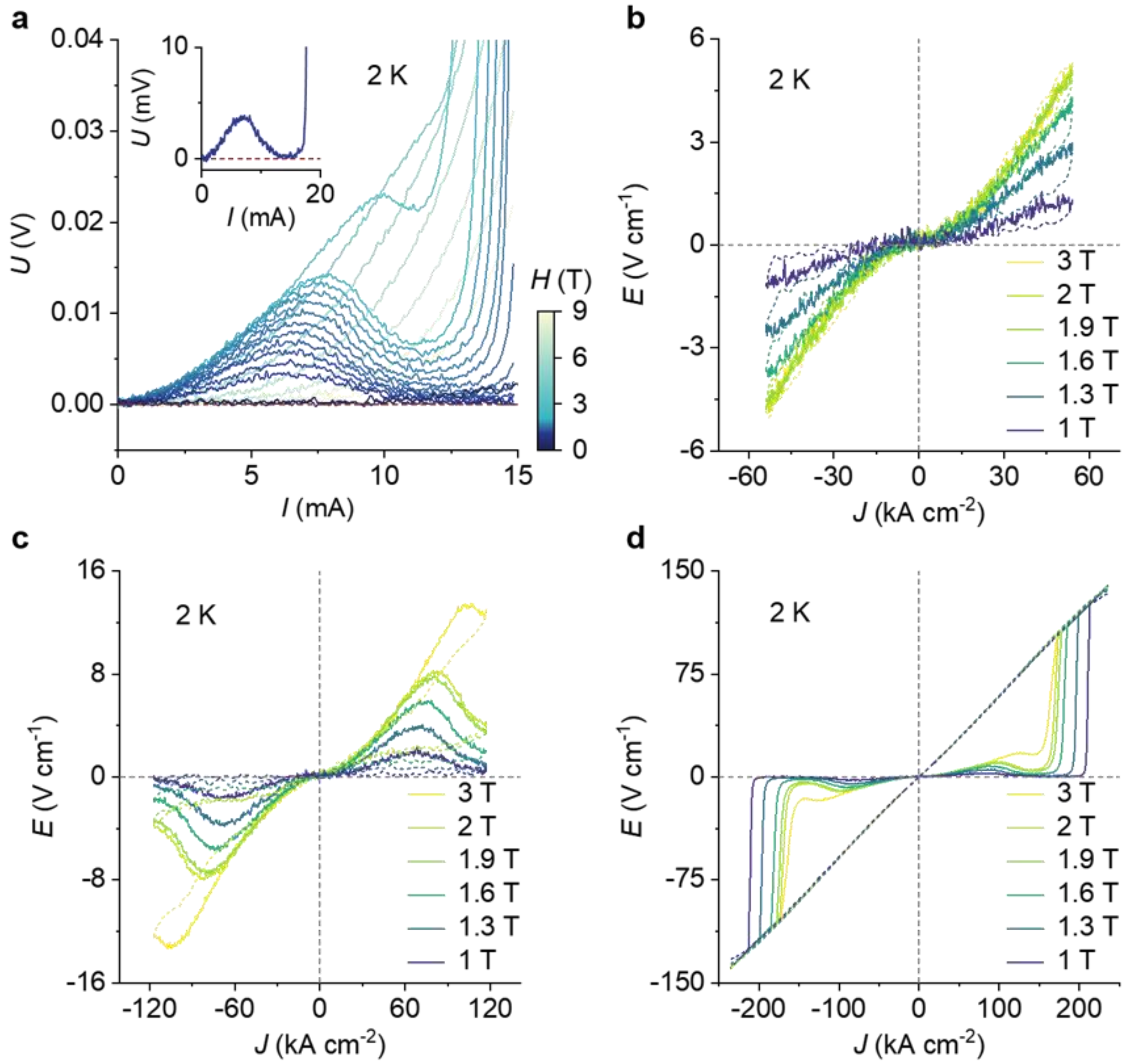

**Extended Data Fig. 3 | Further results for inverse superconducting transition driven by current density. a,** Current-voltage characteristics of overdoped $Eu_xNd_{1-x}NiO_2$ measured at 2 K under magnetic fields up to 9 T. At intermediate fields, increasing current first drives the sample into a zero-voltage superconducting state and subsequently suppresses superconductivity at higher current. Inset, magnified view of low-current region, showcasing current-triggered superconducting states with robust zero-resistance. **b-d,** Closed-loop relationships between current density (*J*) and electric field (*E*) measured over ranges of b, ±60 kA cm$^{-2}$; c, ±120 kA cm$^{-2}$; and d, ±200 kA cm$^{-2}$, giving rise to three characteristic hysteresis loops. 1) Within low current density ranges (e.g., ±60 kA cm$^{-2}$) that are insufficient to trigger the inverse superconducting transitions, the forward and backward current sweeps result in ohmic *J*-*E* relations with approximate superposition and limited mouldability by magnetic-field. 2) Within intermediate current density ranges (e.g., ±120 kA cm$^{-2}$) triggering the inverse superconducting transitions but no quenching, *E* first increased with *J* and abruptly dropped to zero along the forward sweep, and remained at zero along the backward sweep. 3) Within high current density ranges (e.g., ±200 kA cm$^{-2}$) triggering the inverse superconducting transitions and further quenching, *E* increased firstly with *J*, followed by a drop to zero, and then increased abruptly again along the forward sweep; the backward sweep retained ohmic *J*-*E* relations.